\documentclass[a4paper]{jpconf}
\pdfoutput=1  % use this for pdflatex builds, arxiv. For lualatex, comment this

\usepackage{graphicx}   % figures
\usepackage{amsmath}    % math typesetting
\usepackage{microtype}  % micro-typesetting adjustments
\usepackage{lipsum}

\usepackage{hyperref}
\hypersetup{
    hidelinks,
    pdftitle={Simulating the radio emission of dark matter},
    pdfauthor={Michael Sarkis},
    }

\renewcommand{\d}{\mathrm{d}}

\newcommand{\pdiff}[2]{\dfrac{\partial #1}{\partial #2}}

\renewcommand{\tr}{\tilde{r}}
\newcommand{\tE}{\tilde{E}}

\begin{document}
\title{Simulating the radio emission of dark matter for new high-resolution observations with MeerKAT} 

\author{M Sarkis and G Beck}

\address{School of Physics, University of the Witwatersrand, Private Bag 3, WITS-2050, Johannesburg, South Africa}

\ead{michael.sarkis@students.wits.ac.za}

\begin{abstract}
	Recent work has shown that searches for diffuse radio emission by MeerKAT - and eventually the SKA - are well suited to provide some of the strongest constraints yet on dark matter annihilations. To make full use of the observations by these facilities, accurate simulations of the expected dark matter abundance and diffusion mechanisms in these astrophysical objects are required. However, because of the computational costs involved, various mathematical and numerical techniques have been developed to perform the calculations in a feasible manner. Here we provide the first quantitative comparison between methods that are commonly used in the literature, and outline the applicability of each one in various simulation scenarios. These considerations are becoming ever more important as the hunt for dark matter continues into a new era of precision radio observations.
\end{abstract}

% ========================================================================================
% End Title Info
% ========================================================================================
\vspace*{-1em}
\section{Introduction}

Despite decades of work, indirect Dark Matter (DM) searches -- those that look for emission from the annihilation and decay products of DM particles -- are yet to find a signal that can be solely attributed to DM. Until such a detection is made, and as our observing capabilities improve with newer and more sophisticated telescopes, we continue to methodically move through the parameter spaces of candidate DM models and eliminate those that conflict with the data. The recent public release of the MeerKAT Galaxy Cluster Legacy Survey data~\cite{knowlesMeerKATGalaxyCluster2022}, together with recent studies that show the competitiveness of using DM radio emission for indirect detection~\cite{beckRadioFrequencySearchesDark2019,regisEMUViewLarge2021,chanRadioConstraintsDark2021}, provides strong motivation for a renewed and continued effort in radio DM searches. In this work we take a brief but detailed look at the various theoretical aspects involved in the modelling of the radio emission from DM, and comment on how the choice of model will likely play an important role in indirect searches with high-resolution instruments. 

Our analysis includes simulations of the DM host environments for two source targets, the Coma galaxy cluster and the M31 galaxy, and a calculation of the synchrotron emission resulting from the annihilation of Weakly Interacting Massive Particles (WIMPs) therein. We model our DM halos with a set of reasonable source parameters and find the emission after solving the electron propagation equation in each environment. The methods of solving this equation are a major focus point of this work, as the choice of technique used can lead to a non-negligible change in the observed emission, particularly in smaller source targets where diffusion effects are significant. With $< 10$ arcsecond resolution capabilities, observations with MeerKAT (and soon the SKA) are for the first time able to probe the inner regions of these targets, which is where the strongest constraints on DM can be found. Therefore, accurate spatial modelling of these targets is essential for us to make full use of the new data. 

\section{Modelling}\label{sec:modelling}

The two source targets in this work, the Coma galaxy cluster and the M31 galaxy, were chosen for their well-characterised properties in the literature. Of particular importance are the profiles of their magnetic fields and thermal gas densities; as these quantities appear in the modelling process (but are often underspecified), the uncertainty of the final solution depends strongly on the treatment of these factors~\cite{sarkisDiffusing2022}. However, since the simulation of the halo environment is not the central focus of this work (and for the sake of brevity), we refer the reader to the following sources for details regarding the parameters in the Coma cluster~\cite{bonafedeComaClusterMagnetic2010,lokasDarkMatterDistribution2003} and in the M31 galaxy~\cite{ruiz-granadosMAGNETICFIELDSOUTER2010,tammStellarMassMap2012}. 

In each halo environment, the emission of synchrotron radiation will be determined by the spatial and energy equilibrium distribution of charged annihilation products, $\psi(\mathbf{x},E)$. In this work the products considered are electrons and positrons. The evolution of these distributions over time is then given by the following propagation equation, which includes the dominant effects of energy losses and spatial diffusion:
\begin{equation}\label{eqn:propagation}
    \pdiff{\psi(\mathbf{x},E)}{t} = \nabla\cdot\left[D(\mathbf{x},E)\nabla\psi(\mathbf{x},E)\right] + \pdiff{}{E}\left[b(\mathbf{x},E)\psi(\mathbf{x},E)\right] + Q(\mathbf{x},E) .
\end{equation}
Here $D$, $b$ and $Q$ are the diffusion, energy-loss and DM annihilation source functions respectively, and the determination of the exact forms of these functions follows the methods laid out in~\cite{sarkisDiffusing2022}.

\subsection{Solving the propagation equation}\label{sec:methods}

We determine the equilibrium electron distribution $\psi$ using two independent techniques. The first, referred to here as the `Green's  Function (GF) method'~\cite{beckRadioFrequencySearchesDark2019,colafrancescoMultifrequencyAnalysisNeutralino2006}, uses a Green's function with simplified forms of $D$ and $b$ to solve Eq.~\ref{eqn:propagation} semi-analytically. The second, referred to as the `Alternating Direction Implicit (ADI) method'~\cite{strongPropagationCosmicRay1998,regisLocalGroupDSph2015}, uses a numerical approach to solve Eq.~\ref{eqn:propagation} iteratively. In both methods we consider the halo environment to be spherically symmetric, so that $\mathbf{x}$ may be replaced by $r$ in Eq.~\ref{eqn:propagation}. We also note here that we have assumed a simplified form of $D$, which would be a tensor in a more general case. As our methodology closely follows the above-mentioned literature, we only summarise these methods and point out any major differences in the following sections.

\subsubsection*{GF method}

If the forms of the diffusion and energy-loss functions are simplified so that they have no spatial dependence, a solution to Eq.~\ref{eqn:propagation} can be found directly with the use of Green's functions and image charges. However, these simplifications often have an impact on the calculated emission (for a review on this topic, see~\cite{sarkisDiffusing2022}). In this work we use non-weighted averages for the magnetic field and thermal gas densities, found using an averaging scale radius that matches the scale radius of the DM halo. This choice encapsulates the region in the halo that contains the majority of WIMP annihilations -- and thus best represents the spatial structure of the halo -- while allowing us to forgo any explicit spatial dependence in Eq.~\ref{eqn:propagation}. Now, the equilibrium distribution of electrons in the halo can be calculated using 
\begin{equation}\label{eqn:greens}
	\psi(r,E) = \dfrac{1}{b(E)} \int^{m_{\chi}}_E \mathrm{d}E' G(r,\Delta v) Q(r,E') \,,
\end{equation}	
with $m_{\chi}$ as the WIMP mass and the Green's function ($G$) given by 
\begin{equation}
	G(r,\Delta v) = \dfrac{1}{\sqrt{4\pi\Delta v}} \sum_{n=-\infty}^{\infty}(-1)^n \int_0^{r_{\textrm{max}}} \d r' \dfrac{r'}{r_n}\left[\exp{\left(-\frac{(r'-r_n)^2}{4\Delta v}\right)} - \exp{\left(-\frac{(r'+r_n)^2}{4\Delta v}\right)} \right]\dfrac{Q(r')}{Q(r)} .
\end{equation}
Here $r_{\mathrm{max}}$ is the maximum radius for any diffusion processes and $r_n = (-1)^n r + 2nr_{\mathrm{max}}$ is the location of the $n^{th}$ image charge. The quantity $\Delta v$ is calculated as 
\begin{equation}
	\Delta v = v(E) - v(E')\, ,
\end{equation}
where
\begin{equation}
	v(E) = \int_E^{m_{\chi}} \d x\, \dfrac{D(x)}{b(x)} \,.
\end{equation}
% \begin{equation}
% 	v(u(E)) = \int_{u_{\mathrm{min}}}^{u(E)} \d x D(x) \qquad \mathrm{and} \qquad u(E) = \int_E^{E_{\mathrm{max}}} \dfrac{\d x}{b(x)} \;.
% \end{equation}

\subsubsection*{ADI method}

In this method, we discretise Eq.~\ref{eqn:propagation} and solve for the equilibrium distribution iteratively. Since the ADI method retains the radial dependence in the diffusion and energy loss functions (where the GF method does not), the problem becomes 2-dimensional in energy and space. Using a traditional finite-difference technique in this scenario could be computationally expensive, which is why we opt for a method that uses so-called `operator splitting' to treat each dimension separately and divide the problem into smaller, more manageable pieces. Thus, during each step of the method, we use a general form of the 1-dimensional Crank-Nicolson (CN), scheme (see, for instance,~\cite{pressNumericalRecipesArt2007}) which is a finite-differencing technique that includes the average of second-order implicit and explicit terms in the updating equation, thereby leveraging the unconditional stability of a fully implicit method while maintaining second-order accuracy in space and time. This scheme is relatively easy to solve, as the updating equation turns out to be a set of linear equations with tridiagonal coefficient matrices. We write this, as in~\cite{strongPropagationCosmicRay1998,regisLocalGroupDSph2015}, as
\begin{equation}\label{eqn:cn_isolated}
	-\dfrac{\alpha_1}{2}\psi^{n+1}_{x-1} + \left(1+\dfrac{\alpha_2}{2}\right)\psi^{n+1}_{x} - \dfrac{\alpha_3}{2}\psi^{n+1}_{x+1} = \dfrac{\alpha_1}{2}\psi^{n}_{x-1} + \left(1- \dfrac{\alpha_2}{2}\right)\psi^{n}_{x} + \dfrac{\alpha_3}{2}\psi^{n}_{x+1} + Q_x\Delta t .
\end{equation}
Here $n$ is the temporal grid index (with the spacing between indices given by $\Delta t$) and $x$ represents either the energy or spatial grid index. The forms of the $\alpha$ coefficients, which encapsulate the diffusion and energy loss effects, need to be found by discretising the relevant operators from Eq.~\ref{eqn:propagation}. The scheme we have used for this is as follows:
\begin{multline}\label{eqn:rdiff}
	\dfrac{1}{r^2}\pdiff{}{r}\left(r^2D\pdiff{\psi}{r}\right) \xrightarrow[\text{discretisation}]{} C_{\tr}^{-2} \left[\dfrac{\psi_{i+1}-\psi_{i-1}}{2\Delta \tr}\left.\left(\log(10)D	+ \pdiff{D}{\tr}\right)\right\vert_{\tr = \tr_i} \right.\\
	+ \left.\dfrac{\psi_{i+1}-2\psi_{i}+ \psi_{i-1}}{\Delta \tr^2}\left.D\right\vert_{\tr = \tr_i} \right]
\end{multline} 
for the radial operator and 
\begin{equation}\label{eqn:Ediff}
	\pdiff{}{E}(b\psi) \xrightarrow[\text{discretisation}]{} C_{\tE}^{-1}\left[\dfrac{b\vert_{\tE = \tE_j} (\psi_{j+1}-\psi_{j})}{\Delta \tE}\right]
\end{equation}
for energy operator, where $C_{\tr}=(r_0\log(10)10^{\tr_i})$, $C_{\tE} = (E_0\log(10)10^{\tE_j})$ and $\Delta \tr$, $\Delta \tE$ represent the radial and energy grid spacings, respectively. We use $i$ and $j$ to denote positions in the radial and energy grids, and have omitted the temporal indices as these forms will apply to both implicit and explicit terms in the same way. The vertical bars denote that the functions which they are attached to are evaluated at the given grid index. We have also made the variable transformations $\tr = \log_{10}(r/r_0)$ and $\tE = \log_{10}(E/E_0)$ (similarly to~\cite{regisLocalGroupDSph2015}, except with base 10 instead of $\e$), which allows us to more accurately track the electron distribution in our grids when the involved processes operate over a wide range of physical scales. Finally, note that in the case of energy losses, we only consider upstream differencing. 

The $\alpha$ values can now be found by taking Eqs.~\ref{eqn:rdiff} and~\ref{eqn:Ediff} and equating coefficients with Eq.~\ref{eqn:cn_isolated}; once these are found, the updating equation can be solved with some matrix solution algorithm. If we represent the discretisation schemes shown above by the symbol $\Psi$, the overall iterative solution can be summarised with the steps
\begin{align}
    \psi^{n+1/2} &= \Psi_{\tE}(\psi^{n}) \, \nonumber \\
    \psi^{n+1} &= \Psi_{\tr}(\psi^{n+1/2}) \, ,
\end{align}
which are repeatedly solved (using Eq.~\ref{eqn:cn_isolated}) until the value of $\psi$ has converged to the equilibrium value. The other minutiae of this method, including initial and boundary conditions, convergence criteria and stability considerations, can be found in~\cite{strongPropagationCosmicRay1998,regisLocalGroupDSph2015}. 

\subsection{Synchrotron emission}

Once found via the GF or ADI methods, the equilibrium distribution is used to calculate the radio emissivity, given by
\begin{equation}\label{eqn:emm}
j_{\mathrm{sync}} (\nu,r) = \int_{0}^{m_\chi} \d E \, \psi_{e^{\pm}}(E,r) P_{\mathrm{sync}} (\nu,E,r) \; ,
\end{equation}
where $\nu$ is the synchrotron frequency, $\psi_{e^{\pm}}$ is the sum of electron and positron equilibrium distributions and $P_{\mathrm{sync}}$ is the power emitted by an electron with an energy of $E$ (this is calculated as in~\cite{beckRadioFrequencySearchesDark2019}). The emissivity is then used to calculate the two main results in this work. Firstly, the azimuthally averaged surface brightness curves,
\begin{equation}\label{eqn:sb}
I_{\mathrm{sync}} (\nu,r, \Theta, \Delta \Omega) = \int_{\Delta\Omega}\d \Omega \int_{l.o.s.} \d l \, \frac{j_{\mathrm{sync}}(\nu,l)}{4 \pi} \; , 
\end{equation}
where $l.o.s.$ is the line-of-sight to a point in the halo at radius $r$, which makes an angle of $\Theta$ with the centre of the halo, and $\Delta \Omega$ is the solid angle over which the surface brightness is calculated. In this work we show results for a single representative frequency of $\nu = 0.5$ GHz. Secondly, we calculate the integrated flux density by
\begin{equation}\label{eqn:flux}
S_{\mathrm{sync}} (\nu,R) = \int_0^R \d^3r^{\prime} \, \frac{j_{\mathrm{sync}}(\nu,r^{\prime})}{4 \pi d_L^2} \; ,
\end{equation}
where the emissivity is integrated over the region enclosed by $R$ and $d_L$ is the luminosity distance to the target. For the results shown in this work we consider $R$ to be the virial radius of the halo.

\section{Results}\label{sec:results}

\begin{figure}[h!]
	\centering
	\hspace*{-3.4em}%
    \includegraphics[height=5.75cm]{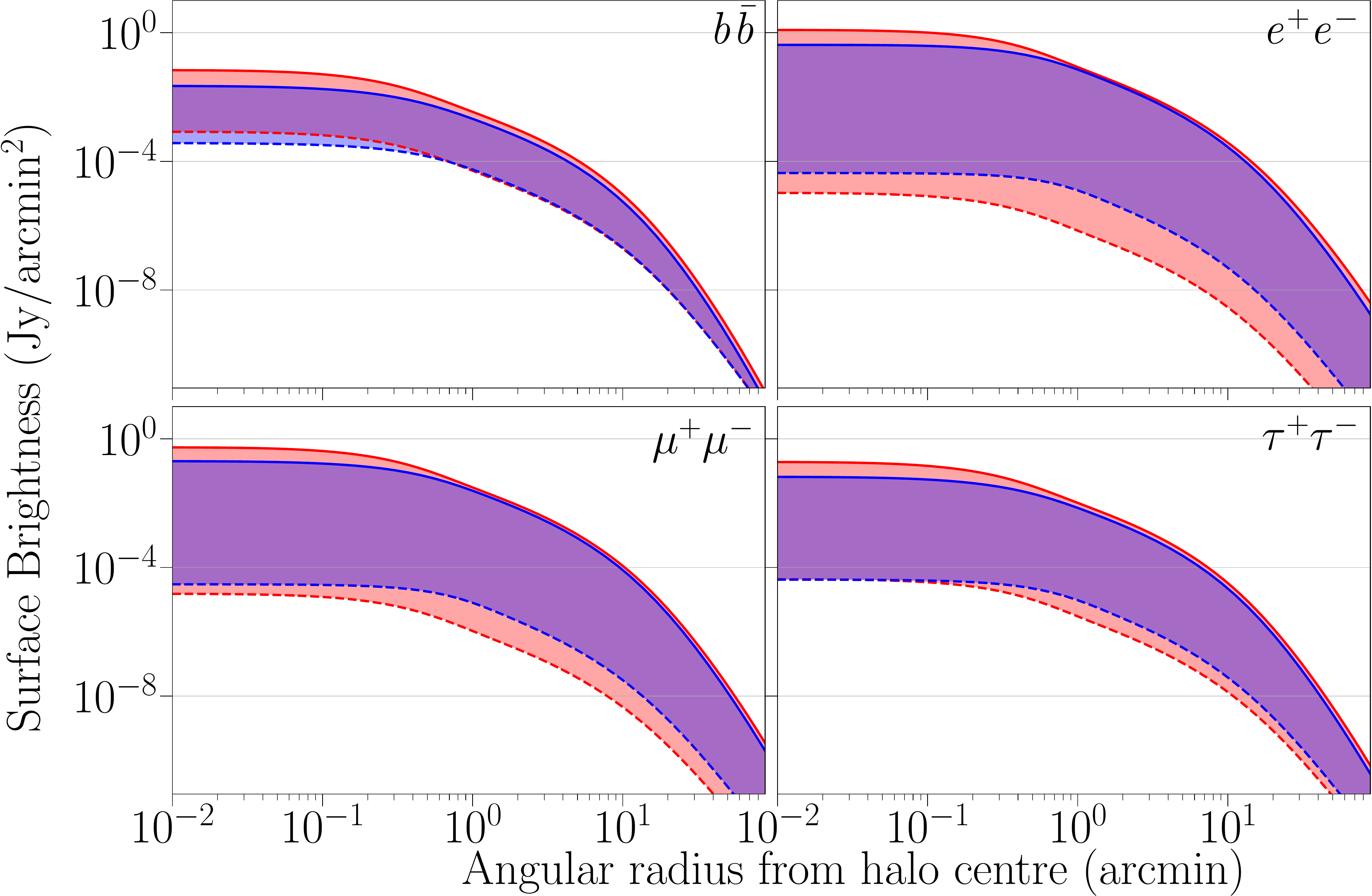}%
	\hspace*{0.5em}%
    \includegraphics[height=5.75cm]{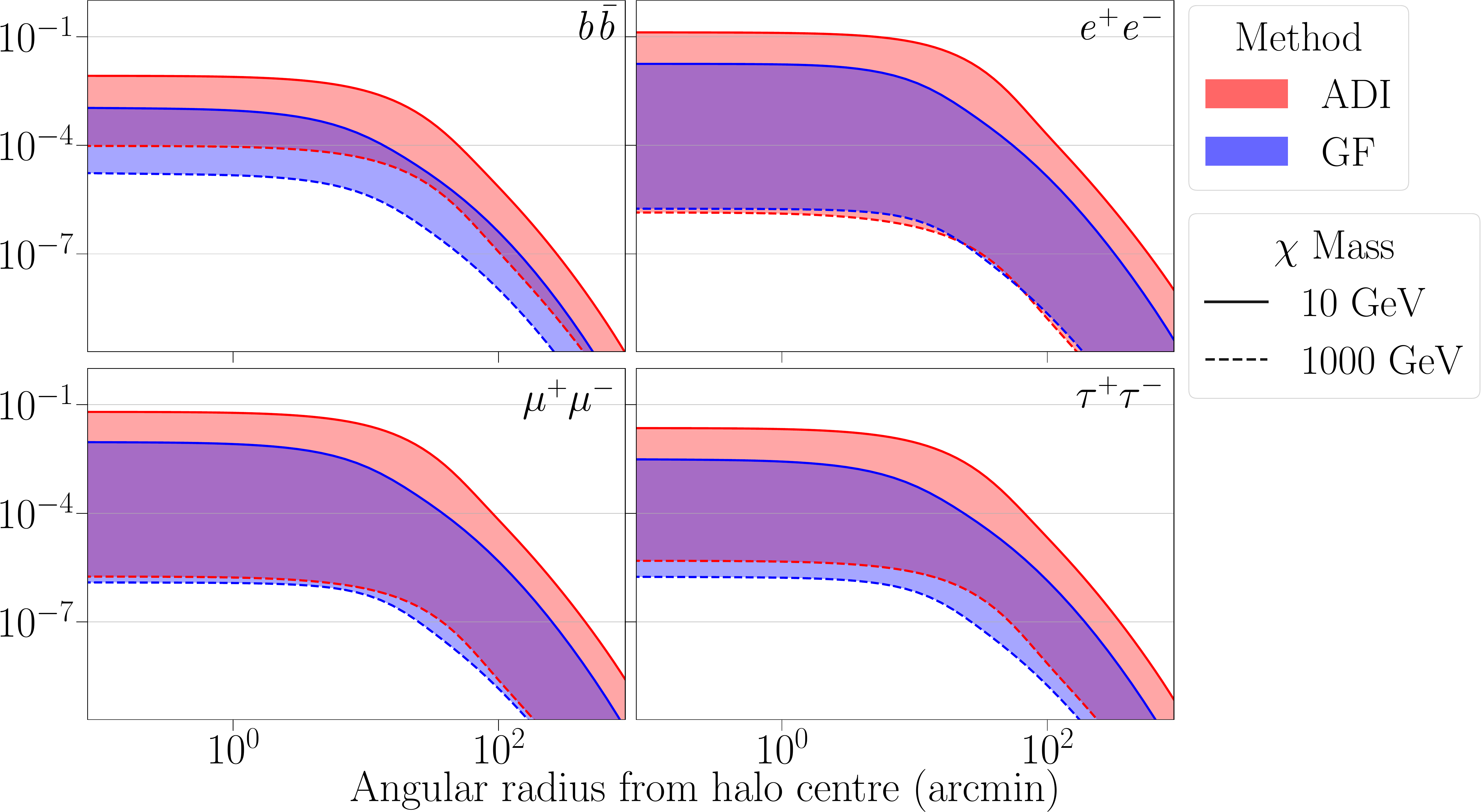}%
	\caption{Surface brightness curves for the Coma galaxy cluster (left-hand panels) and the M31 galaxy (right-hand panels). Each of the four panels show different annihilation channels, given by the label in the top right of each plot. The ADI and GF methods are represented by the red and blue colours respectively, and the region in which the results overlap are given by the combination of these (the purple colour). These shaded regions represent the full mass range of the WIMPs (from 10 to 1000 GeV), and the domain of each panel runs up until the halo's virial radius $R$ (in angular units).}\label{fig:sbs}
\end{figure}

\begin{figure}
	\centering
	% \hspace*{-2.75em}%
    \includegraphics[height=5.25cm]{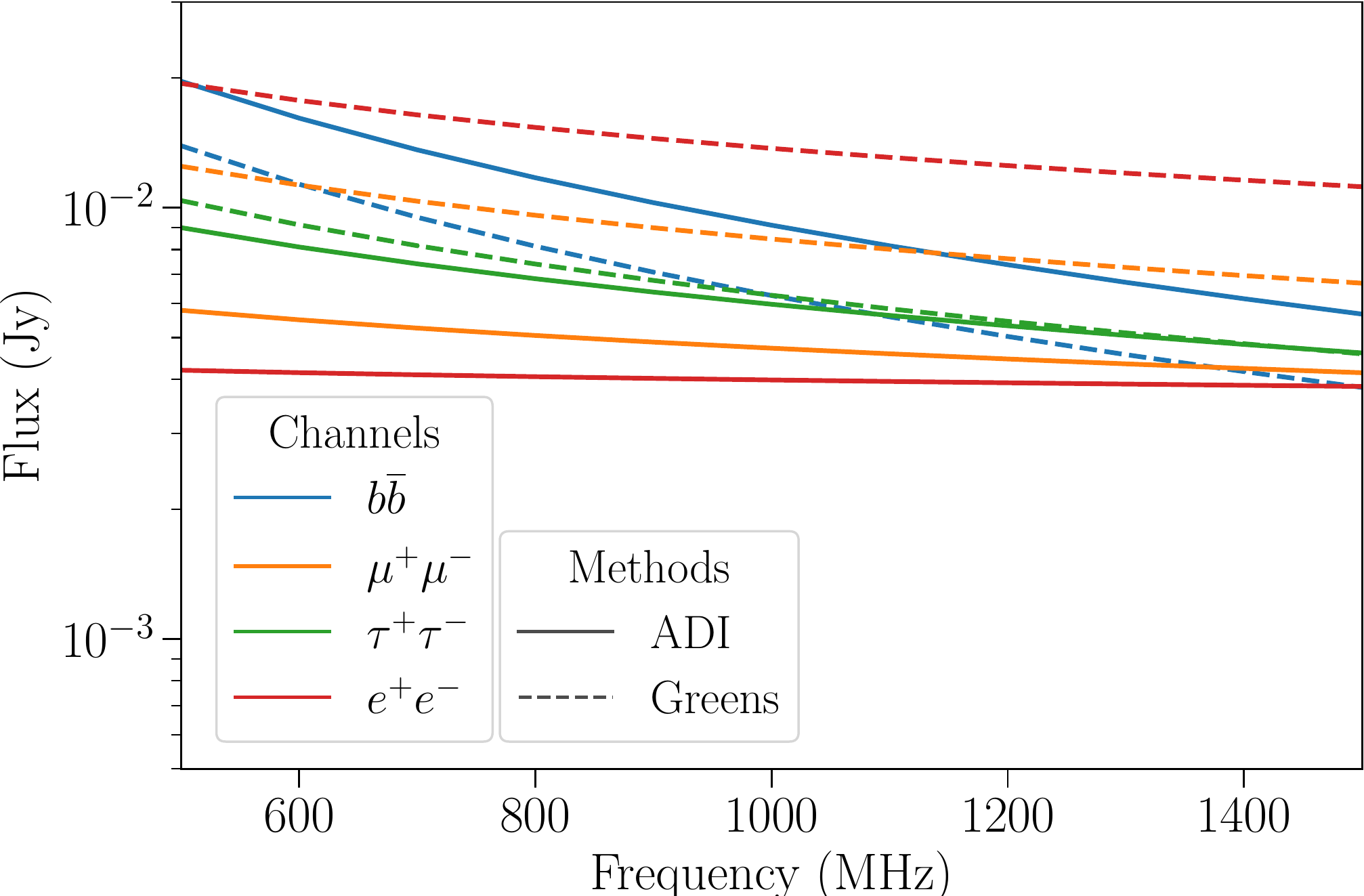}%
	\hspace*{0.5em}%
    \includegraphics[height=5.25cm]{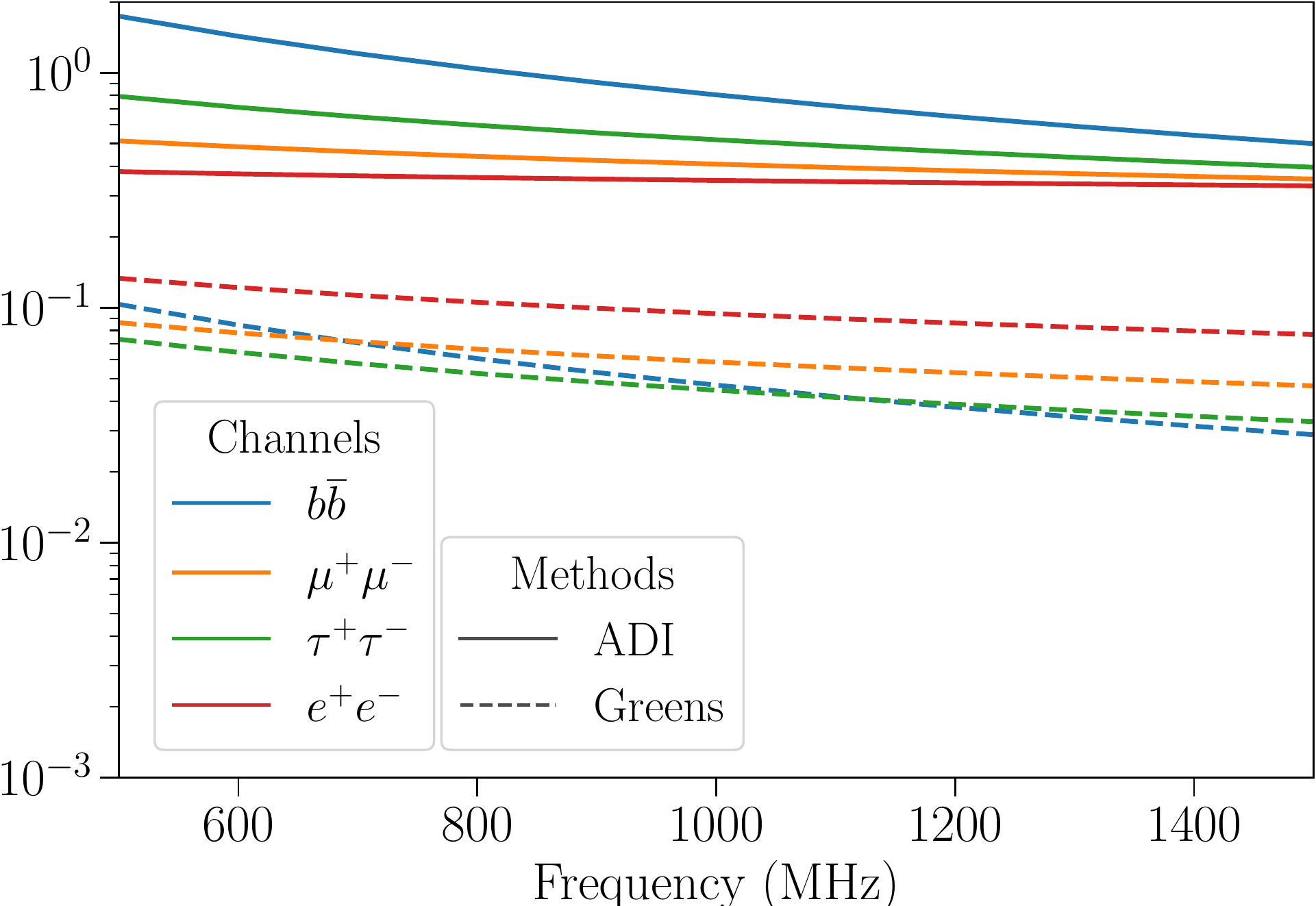}
	\caption{The integrated fluxes, calculated using Eq.~\ref{eqn:flux}, for the Coma galaxy cluster (left) and the M31 galaxy (right). The different linestyles represent the two solution methods presented in Sec.~\ref{sec:methods}, and each colour indicates the use of a different annihilation channel.}\label{fig:fluxes}
\end{figure}

Here we provide the details of the simulations we have performed, and show the results for two observables: the radio surface brightness (Eq.~\ref{eqn:sb}) and integrated flux (Eq.~\ref{eqn:flux}). We use a set of reasonable source parameters for the halo environments that respect observational constraints, and aim to use WIMP parameter values that are representative of the many viable candidates. We thus consider a large range of particle masses, from 10 to 1000 GeV, and use a set of four annihilation channels, $\{b\overline{b}, e^+e^-, \mu^+\mu^-, \tau^+\tau^-\}$. Since the focus of this work is on a comparison between the two solution methods, particurly in the way that they differ with various source targets, we show the results side-by-side and in the same manner for both targets. In Fig.~\ref{fig:sbs} we show the surface brightness curves for the Coma cluster (left-hand panels) and M31 (right-hand panels), and Fig.~\ref{fig:fluxes} shows the integrated fluxes from the same targets for a range of frequencies. 

\section{Discussion and conclusions}\label{sec:discussion}
In Figs.~\ref{fig:sbs}, we see generally good agreement between the GF and ADI methods, which can be inferred from the significant amount of overlap between the curves in each panel. Noticeably however, we see more disagreement (less overlap) between the methods in the M31 galaxy than we do for the Coma galaxy cluster. Our explanation for this lies in the mathematical techniques employed by each method, and how they each treat the spatial dependence of the diffusion function in particular. In the galaxy cluster environment of Coma, diffusion effects are negligible on sufficiently large scales~\cite{colafrancescoMultifrequencyAnalysisNeutralino2006,sarkisDiffusing2022}, whereas in the physically smaller galaxy, diffusion effects start to influence the surface brightness distribution at all relevant scales. Since the GF and ADI methods leverage a spatially independent and dependent diffusion function (respectively), the resulting equilibrium distributions will tend to differ in the environments where the length scales in question do not greatly exceed the diffusion length, as is the case for M31. This trend is also seen in the fluxes from Fig.~\ref{fig:fluxes}, which show a clear disagreement in all channels for the M31 galaxy, and relative agreement in all channels in the Coma cluster. Based on these results and the comparison of target environments presented in~\cite{sarkisDiffusing2022}, we also expect that smaller target environments (like the dwarf spheroidal satellite galaxies of the Milky Way) would show further disagreement between the solution methods, as diffusion effects would be even more significant in these environments. 

The other notable result we see from these simulations is that the methods differ on small scales, even in the large Coma cluster. This is significant, as the inner regions of the DM halos are where we would observe the strongest emission. With high-resolution radio interferometers allowing us to resolve these smaller scales, our models could be tested against the strongest possible DM emission, allowing us to find more stringent constraints on DM properties than previously possible. In this regard, the surface brightness curves displayed here would be especially valuable results when determining new observational limits, as their emission profiles are highly dependent on the spatial structure of the DM halo. 

With the impressive spatial resolution of telescopes like MeerKAT and the SKA, we are now able to probe the inner regions of these DM halos -- regions which have formerly been hidden from our view. The need for accurate modelling techniques is thus more necessary than ever before, and the considerations presented in this work should help inform the modelling choices made in future radio searches for DM.

\ack
This work is based on the research that was supported by the National Research Foundation of South Africa (Bursary No. 112332). G.B. acknowledges support from a National Research Foundation of South Africa Thuthuka grant no. 117969.

% ========================================================================================
% End Final Section
% ========================================================================================
\section*{References}
\bibliography{references/saip2022}

\end{document}